\documentclass[11pt]{article}

\usepackage{fullpage, hyperref,bbm,amssymb,amsfonts,amsthm}
\usepackage[noend]{algorithmic}
\usepackage{algorithm}
\usepackage{algorithm}
\usepackage{setspace}

\newcommand{\<}{\langle}
\renewcommand{\>}{\rangle}

\newcommand{\cZ}{\mathcal{Z}}

\renewcommand{\sim}{\mathrm{sim}}
\renewcommand{\prec}{\mathrm{prec}}
\newcommand{\fail}{\mathrm{fail}}
\newcommand{\median}{\mathrm{median}}
\newcommand{\anc}{\mathrm{anc}}

\newcommand{\tr}{\mathrm{tr}}

\newtheorem*{theorem}{Theorem}
\newtheorem{lemma}{Lemma}

\newtheorem*{problem}{Problem}

\begin{document}
\title{Quantum Algorithm for Preparing Thermal Gibbs States -- Detailed Analysis} 
\author{
Chen-Fu Chiang\thanks{School of Electrical Engineering
and Computer Science,
University of Central Florida, Orlando, FL~32816, USA. Email:
\texttt{cchiang@eecs.ucf.edu}} \quad
Pawel Wocjan\thanks{School of Electrical Engineering and Computer Science,
University of Central Florida, Orlando, FL~32816, USA. Email:
\texttt{wocjan@eecs.ucf.edu}}
}		
		
\maketitle

\begin{abstract}
In a recent work \cite{PW:09}, {\sc Poulin} and one of us presented a quantum algorithm
for preparing thermal Gibbs states of interacting quantum systems.  This algorithm is based on Grover's technique for quantum state engineering, and its running time is dominated by the factor $\sqrt{D/\cZ_\beta}$, where $D$ and $\cZ_\beta$ denote the dimension of the quantum system and its partition function at inverse temperature $\beta$, respectively.

We present here a modified algorithm and a more detailed analysis of the errors that arise due to imperfect simulation of Hamiltonian time evolutions and limited performance of phase estimation (finite accuracy and nonzero probability of failure).  This modification together with the tighter analysis allows us to prove a better running time by the effect of these sources of error on the overall complexity.  We think that the ideas underlying of our new analysis could also be used to prove a better performance of quantum Metropolis sampling by {\sc Temme} et al. \cite{TOVPV:09}.
\end{abstract}

\section{Introduction}
The ability to efficiently prepare thermal Gibbs states of arbitrary quantum systems at arbitrary temperatures on a quantum computer would lead 
to a multitude of applications in condensed matter, quantum chemistry and high energy physics \cite{PW:09, TOVPV:09, TD:00}.  For example, we could estimate partition and correlation functions of fermionic and frustrated sytems.  For these systems, the approach of first applying the ``quantum-to classical map'' \cite{Suzuki:88} and then using the classical Monte Carlo method fails because the mapping does not conserve the positivity of statistical weights.

We consider an arbitrary Hamiltonian $H$ with spectral decomposition
\begin{equation}
H = \sum_{a=1}^D E_a|\psi_a\>\<\psi_a|\,.
\end{equation} 
The {\em thermal Gibbs state} of the system at inverse temperature
$\beta$ is given by
\begin{equation}
\rho_\beta := \sum_a \frac{e^{-\beta E_a}}{\cZ_\beta} |\psi_a\>\<\psi_a|
\end{equation}
where $\cZ_\beta:=\sum_a e^{-\beta E_a}$ denotes the partition function.
			
The formal definition of preparing thermal Gibbs states is as follows:
\begin{problem}[Thermalizing quantum states]
Let $H$ be a Hamiltonian, $\beta$ an inverse temperature
and $\epsilon \in (0,1)$ a parameter describing the desired accuracy. We 
consider the problem to prepare a state $\tilde{\rho}_\beta$ that is 
$\epsilon$-close to	the thermal Gibbs state $\rho_\beta$ with respect to
trace distance\footnote{Recall that the trace distance is defined to be $\frac{1}{2}\tr \sqrt{XX^\dagger}$, where $X=\rho_\beta - \tilde{\rho}_\beta$.}, i.e., 
\begin{equation}\label{eq:errcriteria}
\| \rho_\beta - \tilde{\rho}_\beta \|_{\mathrm{tr}} \le
\epsilon.
\end{equation}
We refer to the process of preparing such state as thermalizing the quantum
system. We seek to determine efficient quantum circuits that realize such 
thermalizing process. 
\end{problem}

We assume that the energies satisfy $E_a \in [0,\frac{\pi}{4}]$.  If we initially only know that the spectrum of the Hamiltonian $H$ is contained in the interval $[\ell,u]$, then the shifted and rescaled Hamiltonian $4(H - \ell I)/(\pi (u-\ell))$ satisfies the condition of this assumption.  The thermal Gibbs state is invariant under shifting of the spectrum.  Thus, we have to rescale the inverse temperature by multiplying it by $u-\ell$ when working with the new Hamiltonian.
	
There are two types of quantum algorithms for preparing thermal Gibbs states.  The first is a generalization of the Metropolis algorithm.  The Metropolis algorithm \cite{Metropolis:53} can be applied to the special case of classical systems, i.e., systems whose Hamiltonian $H=\sum_a E_a |a\>\<a|$ is diagonal in the computational basis. It offers great flexibility for constructing Markov chains whose limiting distributions  are equal to the desired thermal Gibbs distributions.  The number of times we have to apply the Markov chain scales like $1/\delta$, where $\delta$ is its spectral gap.  Bounding the spectral gap from below for arbitrary systems and neighborhood structures is very difficult.  However, it is possible to prove that the gap is sufficiently large for many practically relevant cases.  

Recently, {\sc Temme} et al. presented an extention of the Metropolis algorithm to quantum sytems \cite{TOVPV:09}.  Their quantum Metropolis sampling makes it possible to implement quantum maps such that their fixpoints are approximately equal to the desired thermal Gibbs states.  Analogously to classical case, the number of time we have to apply the quantum map depends on its spectral gap.  The difficulty of bounding the gap from below remains for general systems and neighborhood structures.  However, numerical experiments in \cite{TOVPV:09} show that the gap scales like $1/N$ for the spin-chain Hamiltonian $H=\sum_k X_k X_{k+1} + Y_k Y_{k+1} + g Z_k$ on $N$ spins.

The second type of algorithm is due to {\sc Poulin} and one of us \cite{PW:09}.  This algorithm behaves like a Las Vegas algorithm, i.e., it always produces a correct output $\tilde{\rho}$ satisfying the requirements of the problem definition.  The time it takes this algorithm to terminate is a random variable.  However, we can bound the expected value.  It is dominated by the factor $\sqrt{D/\cZ_\beta}$.  This square root term occurs because this algorithm is based on an extension of Grover's state engineering technique.  

Once the Grover sampling has terminated we know that we have prepared a state that is close to the desired thermal Gibbs state.  In contrast, we can only guarantee that quantum Metropolis sampling yields a good approximation if we have a lower bound on the spectral gap. But, of course, quantum Metropolis sampling has the potential to outperform the Grover sampling for certain quantum systems.

We modify this Grover sampling and analyze the errors that arise due to imperfect simulation of Hamiltonian time evolutions and limited performance of phase estimation (finite accuracy and nonzero probability of failure) in more detail.  This modification together with the tighter analysis allows us to prove a better running time.  We show that the expressing the effect of these sources of error on the overall complexity is smaller than in the original algorithm.  We also think that the ideas underlying of our new analysis could also be used to prove a better performance of the above quantum Metropolis sampling.
		
This paper is organized as follows. In section \ref{q-alg} we present the structure of the algorithm. We identify three sources of errors that arise due to (i) imperfect simulation of Hamiltonian time evolution, (ii) limited precision of phase estimation, and	(iii) non-zero failure probability of phase estimation. In section
\ref{Sect:ImpHamSim} and \ref{Sect:PEerror} we analyze how the complexity increases when we seek to keep the errors small. Finally, we make our conclusion in Section \ref{conclusions}.
		
\section{Quantum algorithm -- idealized setting}{\label{q-alg}}

To better explain the intuition behind the quantum algorithm, we first ignore
all sources of error. We assume that the unitary $U=\exp(2\pi iH)$ can be implemented perfectly and
efficiently.  The eigenvalues $E_a$ of $H$ correspond to the eigenphases $E_a$ of $U$, using the convention that the phase of $e^{2\pi i E_a}$ is $E_a$. We assume that phase estimation (PE) makes it possible to perfectly resolve the eigenphases, i.e., there is an efficient quantum circuit mapping $|\psi_a\> \otimes |00\ldots 0\>$ onto $|\psi_a\> \otimes |E_a\>$ (this is the case as long as the energy $E_a$ can be written as binary fractions).  The realistic case is analyzed in detail in the following section.
	
The algorithm prepares a purified Gibbs state of the form
	\begin{equation}\label{pureGibbs}
    |\beta\>  = 
    \sum_{a=1}^D \sqrt{\frac{e^{-\beta E_a}}{\cZ_\beta}}
    \underbrace{|\psi_a\>}_{A} \otimes \underbrace{|\varphi_a\>}_{B} \otimes \underbrace{|E_a\>}_{\textrm{energy}}
    \otimes \underbrace{|0\>}_{\textrm{anc}}.
    \end{equation}
 	The states $|\varphi_a\>$ form an orthonormal basis on the $D$-dimensional subsystem $B$.  The $|E_a\>$ are computational basis states of the energy
  	register, which consists of multiple qubits.  These basis states encode the eigenvalues $E_a$ of the eigenvectors $|\psi_a\>$ of $H$.  The ancilla
  	register consists of a single qubit. We obtain the thermal Gibbs state
  	$\rho_\beta$ from $|\beta\>$ by tracing out the subsystems $B$, energy, and
  	anc (see eqn. (\ref{pureGibbs}))
	\begin{equation}
	\rho_\beta = \tr_{\bar{A}}(|\beta\>\<\beta|)\,,
	\end{equation}
	where we use $\bar{A}$ to denote the collection of the above three subsystems (the complement of $A$).

	The algorithms consists of the following steps:
	
\begin{algorithm}[H]\caption{Thermal Gibbs State Preparation at Inverse Temperature $\beta$}\label{alg:Gibbs}
\textbf{Input:}
Prepare the maximally entangled state $|\nu\> =
    		\frac{1}{\sqrt{D}}\sum_a|a\>|a\>$ on the subsystem $AB$.
    		\\
    		\textbf{Step I:}
    		Run phase estimation of $U$ on the $A$-part of $|\nu\>$. Write the 
    		eigenphase into the energy register. \\
    		\textbf{Step II:}
    		Apply the controlled rotation $R = \sum_E |E\>\<E| \otimes R_E$
    		where 
            \[
            R_E =  
			\left(
			\begin{array}{cc}
			  \sqrt{e^{-\beta E}}   & - \sqrt{1-e^{-\beta E}} \\
			  \sqrt{1-e^{-\beta E}} &  \sqrt{e^{-\beta E}}
			\end{array}
			\right)\,.
            \] 
            
            The control is the energy register and the target is the ancilla
            qubit that is initialized in $|0\>$. \\
            Denote the
            resulting state by $|\Psi\>$.\\
            \textbf{Step III:}
            Use a variant of Grover to project $|\Psi\>$ onto 
            the subspace in which the ancilla qubit is in $|0\>$.  Denote the projector onto this subspace by $\Pi_0$.  The Grover iteration is given by
            \[
            G = (2|\Psi\>	\<\Psi| - I )(I - 2\Pi_0)\,.
            \]
            \textbf{Output:} The density matrix $\rho_\beta$ of final state
            $|\beta\>$ by tracing out $\bar{A}$.
	\end{algorithm}  
	
	Let $V$ be an arbitrary unitary. The maximally entangled state $|\nu\>$ is
	invariant under the action of $V\otimes \bar{V}$, i.e., $(V \otimes \bar{V})|\nu\> = |\nu\>$.  Consequently, we can rewrite $|\nu\>$ as
	\begin{equation} 
   	|\nu\> = \frac{1}{\sqrt{D}} \sum_a |\psi_a\> \otimes |\varphi_a\>
    \end{equation}
	by setting $V = \sum_a |\psi_a\>\<a|$ and $|\varphi_a\> = \bar{V}|a\>$. 
	In step I, we obtain the state
	\begin{equation}
    |\Phi\>	= \frac{1}{\sqrt{D}} \sum_a \Big(|\psi_a\> \otimes
			|\varphi_a\>\Big) \otimes |E_a\>\otimes |0\>.
    \end{equation}
 	In step II, we obtain the state
	\begin{equation}
    |\Psi\> = \frac{1}{\sqrt{D}} \sum_a \Big(|\psi_a\> \otimes
			|\varphi_a\>\Big) \otimes |E_a\>\otimes \Big(\sqrt{e^{-\beta E_a}}|0\> + 
			\sqrt{1-e^{-\beta E_a}}|1\> \Big).
    \end{equation}
	Note that the desired purified Gibbs state $|\beta\>$ is equal to 
	\begin{equation}
    \frac{\Pi_0|\Psi\>}{\|\Pi_0|\Psi\>\|}\,,
  \end{equation}
	where $\|\Pi_0|\Psi\>\| = \sqrt{\frac{\cZ_\beta}{D}}$. We apply the variant of Grover algorithm \cite{BBH:96}, which makes it possible to prepare $|\beta\>$ with an expected number of Grover iterations $O(1/\|\Pi_0|\psi\>\|)$.  It is important that we do not need to know the overlap $\|\Pi_0|\Psi\>\|$.
	This shows that we obtain
	\begin{equation}
    |\beta\> = 
    \sum_a^D \sqrt{\frac{e^{-\beta E_a}}{\cZ_\beta}} \, 
    |\psi_a\> \otimes 
    |\varphi_a\> \otimes 
    |E_a\> \otimes |0\>
    \end{equation}
    in step III.
    
\section{Quantum algorithm}
    
  
\subsection{Analysis of simulation error}\label{Sect:ImpHamSim}
		
The first source of error is the inability to implement $U = \exp(2\pi iH)$ perfectly for general $H$. Using techniques \cite{BACS:07, Lloyd:96, Zalka:98} for simulating Hamiltonian time evolutions, we can only implement a unitary $U_\sim$ with $\|U - U_\sim\|\leq \epsilon_\sim$.  The resources grow inversely with the desired accuracy $\epsilon_\sim$.
	
To bound the error arising from imperfect simulation, we use the following
result, which follows the discussion in \cite[Appendix A]{PW:09}.
	  
\begin{lemma}
Let $H$ be a Hamiltonian whose eigenvalues are contained in the interval $[0,\frac{\pi}{4}]$.  Let $U=exp(2\pi iH)$ and $U_\sim$ be a unitary with $\|U-U_\sim\|\le\epsilon_\sim$.
Then, there exists an effective Hamiltonian $H_\sim$ such that $U_\sim=\exp(2\pi i H_\sim)$ and $\|H-H_\sim\|\le \kappa \epsilon_\sim$ where $\kappa$ is a constant.
\end{lemma}
	
Assume phase estimation could perfectly resolve the eigenphases of $U_\sim$.
Then, if we ran the algorithm using $U_\sim$ instead of $U$, then we would prepare the thermal state with respect to the effective Hamiltonian $H_\sim$ instead of $H$.  Thus, it remains to determine how close the corresponding thermal states are close to each other with respect to trace norm. 
	
\begin{lemma}{\label{lem:SimUniPrec}}
Let $H$ and $H_\sim$ be as above.  Then, the corresponding thermal states
\begin{equation}
\rho         := \frac{\exp(-\beta H)}{\tr(\exp(-\beta H))} \quad\mbox{and}\quad
\rho_\sim    := \frac{\exp(-\beta H_\sim)}{\tr(\exp(-\beta H_\sim))}
\end{equation}
satisfy
\begin{equation}
\|\rho - \rho_\sim \|_\tr \le \frac{\epsilon}{2}
\end{equation}
provided that $\epsilon_\sim \le \epsilon^2/(8\kappa \beta)$.
\end{lemma}

\proof The fidelity of $\rho$ and $\rho_\sim$ is given by
\begin{equation}
F(\rho,\rho_\sim) = \tr \sqrt{\sqrt{\rho} \, \rho_\sim \, \sqrt{\rho}}\,.
\end{equation}
Using \cite[Proposition 4]{FG:99} we bound the trace distance between $\rho$ and $\rho_\sim$ as follows
\begin{equation}
\|\rho - \rho_\sim\|_\tr \leq \sqrt{1 - {F(\rho, \rho_\sim)}^2}\,.
\end{equation}
The analysis in \cite[Appendix C]{PW:09} shows that 
\begin{equation}
F(\rho , \rho_\sim) \ge e^{-\beta \kappa
\epsilon_\sim},
\end{equation}
and thus 
\begin{equation}
\|\rho - \rho_\sim\|_\tr \le \sqrt{1 - e^{-2\beta \kappa \epsilon_\sim}} \le 
\sqrt{2\beta \kappa \epsilon_\sim} \le
\frac{\epsilon}{2}.
\end{equation} 
The rightmost inequality follows from $1+x \leq e^x$ for all $x
\in \mathbb{R}$. \qed \\	
		
From now on, we measure the complexity in terms of how many times we have to  
invoke a controlled version of $U_\sim$. If we wish to determine the complexity in terms 
of elementary gates, we have to look at the simulation technique more closely.
	
\subsection{Analysis of Errors in Phase Estimation}\label{Sect:PEerror}
We now show how to prepare a state $\tilde{\rho}$ such that $\|\rho_\sim - \tilde{\rho}\|_\tr \le \epsilon/2$, implying that
$\|\rho - \tilde{\rho}\|_\tr \le \epsilon$ as desired. We analyze the three phases of the algorithm.   

\subsubsection*{Phase I}
We need to run a special variant of phase estimation
\cite{NWZ:09} of $U_\sim$ on $|\nu\>$.  We briefly explain how it works. 
To avoid new definitions, we use $|\psi_a\>$ and $E_a$ to refer to the eigenvectors and
eigenphases of $U_\sim$, respectively.

The usual phase estimation algorithm consists of the following steps \cite{KLM:07}.  The energy register consits of $n=\lceil \log_2 (1/\epsilon_\prec) \rceil$ qubits.  We apply the Hadamard transform to each of the qubits of the energy register,  the controlled-$U_\sim^{2^j}$ gates (controlled by the $j$th qubit of the energy register) on the $A$-part of $\nu$, and the inverse quantum Fourier transform $F^\dagger$ on the energy register.  We measure the $n$ qubits of the energy register in the computational basis and interpret the outcome $b\in[0,2^n-1]$ as the binary fraction $\hat{E}_b:=b/2^n$, which is a very good estimate for $E_a$.  More precisely, the probability of obtaining the estimate $\hat{E}_b$ is given by 
\begin{equation}
\Pr(E_a,\hat{E}_b)=\frac{1}{2^{2n}} \frac{\sin^2(\pi 2^n(E_a - \hat{E}))}{\sin^2(\pi (E_a - \hat{E}_b))}\,.
\end{equation}
We use $|\hat{E}_b\>$ to denote the compuational basis state $|b\>$, which encodes the energy value $\hat{E}_b$. Let $E_a^\pm$ denote the binary fractions that are closest to $E_a$, where we use the convention $E_a^- \le E_a < E_a^+$.  It follows that the probability of obtaining $E_a^+$ or $E_a^-$ is greater or equal to $\frac{8}{\pi^2}\ge \frac{3}{4}$.  Thus, the probability of failure, i.e., the probability of not obtaining one of the closest $n$-bit fractions, is less than $\frac{1}{4}$.  

To reduce the probability of failure to $\epsilon_\fail$, we repeat this quantum circuit $k=\lceil \log_2(1/\epsilon_\fail) \rceil$ times, each time recording the estimate into a new energy register and adjoin a median register that consists of $n$ qubits.  This yields the state $|\Upsilon\>$
\begin{equation}
\frac{1}{\sqrt{D}}
\sum_a |\psi_a\>|\varphi_a\> \otimes 
\Big(
\sum_{b_1} c_{E_a,\hat{E}_{b_1}}|\hat{E}_{b_1}\>_{\mathrm{energy}} \otimes \ldots \otimes 
\sum_{b_k} c_{E_a,\hat{E}_{b_k}}|\hat{E}_{b_k}\>_{\mathrm{energy}}
\Big)
\otimes |0\ldots 0\>_\median \otimes|0\>_\anc\,,
\end{equation}
where the amplitudes $c_{E_a,\hat{E}_{b_\ell}}$ satisfy $|c_{E_a,\hat{E}_{b_\ell}}|^2 = \Pr(E_a,\hat{E}_{b_\ell})$ for $\ell=1,\ldots,k$.  

The median circuit determines the median of $\hat{E}_{b_1}$, \ldots, $\hat{E}_{b_k}$ and writes it into the median register.  Reordering the registers, we may write the resulting states as 
\begin{equation}
|\tilde{\Upsilon}\> = 
\frac{1}{\sqrt{D}}\sum_a |\psi_a\>|\varphi_a\>\otimes 
\big(
c^{\pm}_a|E_a^{\pm}\>_\median \otimes |\mu_a^{\pm}\>_{\mathrm{energy}^{\otimes n}} + 
|\xi_a\>_{\median\otimes\mathrm{energy}^{\otimes n}} 
\big)
\otimes |0\>_\anc\,.
\end{equation}
where 
\begin{itemize}
\item the states $|\mu_a^{\pm}\>$ are supported only on the states $|\hat{E}_{b_1}\>\otimes\cdots\otimes|\hat{E}_{b_k}\>$ such that the median of $\hat{E}_{b_1},\ldots,\hat{E}_{b_k}$ is equal to $E_a^\pm$, and
\item the states $|\xi_a\>$ are orthogonal to $|E_a^{\pm}\>\otimes|\mu_a^\pm\>$.
\end{itemize}
It follows from the analysis in \cite{NWZ:09, JVV:86} that the amplitudes $c_a^{\pm}$ satisfy
\begin{equation}
1 -\epsilon_\fail \leq |c_a^+|^2 + |c_a^-|^2 \le 1\,,\quad
0 < \| |\xi_a\> \|^2 \leq \epsilon_\fail\,.
\end{equation} 
This means that the probability of the median not being one of the closest binary fractions $E_a^\pm$ to $E_a$ is less than or equal to $\epsilon_\fail$.
The advantage of combining phase estimation with the powering techinque for approximation algorithms is that we only need to invoke a controlled version of $U_\sim$
\begin{equation}\label{eqn:phaseestcomplexity}
\lceil (1/\epsilon_\prec)\log(1/\epsilon_\fail) \rceil
\end{equation}
instead of $O((1/\epsilon_\prec)(1/\epsilon_\fail))$ when using phase estimation alone \cite{NC:00}.

To keep the notation simple, we use $|E_a^{\pm}\>$ to denote the tensor product $|E_a^{\pm}\> \otimes |\mu_a^{\pm}\>$.  Using this convention, we write the state after step I (phase estimation) as 
\begin{equation}
	|\tilde{\Phi}\>	 =   \frac{1}{\sqrt{D}}\sum_a |\psi_a\>|\varphi_a\>\otimes
    (c^{\pm}_a|E_a^{\pm}\> + |\xi_a\>)\otimes |0\>\,.
\end{equation}

\subsubsection*{Phase II}
The $R$-operation is controlled by the energy value contained in the median register. After step II,
$|\tilde{\Phi}\>$ evolves to the state
\begin{eqnarray}
|\tilde{\Psi}\> & = & 
\underbrace{\frac{1}{\sqrt{D}}\sum_a |\psi_a\>|\varphi_a\> \otimes
c_a^{\pm}|E_a^{\pm}\> \otimes \Big(\sqrt{e^{-\beta E_a^{\pm}}}|0\> + \sqrt{1-
e^{-\beta E_a^{\pm}}}|1\>\Big)}_{\mbox{$|\psi\>$}}  + \\
&   & 
\underbrace{\frac{1}{\sqrt{D}}\sum_a |\psi_a\>|\varphi_a\>\otimes
R(|\xi_a\> \otimes |0\>)}_{\mbox{$|\xi\>$}}\,.
\end{eqnarray}
As a consequence of the property $\<\xi|\psi\> = 0$, we have 
\begin{equation}\label{eqn:proberrorbound}
\||\xi\>\|^2 = \frac{1}{D}\sum_a\<\xi_a|\xi_a\> \leq \epsilon_{\fail}
\quad \mbox{ and } \quad 
1 -\epsilon_{\fail} \leq \||\psi\>\|^2 \le 1 .
\end{equation}  

\subsubsection*{Phase III}
Let $|\tilde{\beta}\>$ be the state obtained by applying Grover's algorithm to $|\tilde{\Psi}\>$, i.e., 
\begin{equation}
|\tilde{\beta}\> = \frac{\Pi_0|\tilde{\Psi}\>}{\|\Pi_0|\tilde{\Psi}\>\|}.
\end{equation}
Let $\tilde{\rho}$ be the reduced density operator of $|\tilde{\beta}\>$ over
$\bar{A}$. We need to bound $\|\Pi_0|\tilde{\Psi}\>\|$ from below to obtain an upper bound on the expected number of Grover iterations. 
We also need to show that $\tilde{\rho}$ is close to $\rho_\sim$. This is done in the following lemma. 

\begin{lemma}\label{lem:errorbound}
Let $\tilde{\rho} = \tr_{\bar{A}}(|\tilde{\beta}\>\<\tilde{\beta}|)$. This density operator has the form 
\begin{equation}
\tilde{\rho} =
\frac{\tr_{\bar{A}}{\big(\Pi_0 \big |\psi\>\<\psi|\Pi_0\big)}}{\<\tilde{\Psi}|\Pi_0|\tilde{\Psi}\>} 
+ \frac{\tr_{\bar{A}}{\big(\Pi_0 |\xi\>\<\xi|\Pi_0\big)}}{\<\tilde{\Psi}|\Pi_0|\tilde{\Psi}\>}
\end{equation}
that satisfies 
\begin{equation}\label{eqn:simpeerror}
\|\rho_\sim - \tilde{\rho}\|_\tr
\leq
\|\rho_\sim - \frac{\tr_{\bar{A}}{(\Pi_0(|\psi\>\<\psi|)\Pi_0)}}{\<\tilde{\Psi}|\Pi_0|\tilde{\Psi}\>}\|_\tr +
\|\frac{\tr_{\bar{A}}{(\Pi_0(|\xi\>\<\xi|)\Pi_0)}}{\<\tilde{\Psi}|\Pi_0|\tilde{\Psi}\>}\|_\tr\,
\leq \frac{\epsilon}{4} + \frac{\epsilon}{4} = \frac{\epsilon}{2} \, , 
\end{equation}
provided that $\epsilon_\fail = e^{-\beta}\,\epsilon^2$ and
$\epsilon_\prec= \epsilon/(32\beta).$
\proof
Observe that the off-diagonal terms $\Pi_0(|\psi\>\<\xi| +|\xi\>\<\psi|)\Pi_0$ in 
${\big(\Pi_0|\tilde{\Psi}\>\<\tilde{\Psi}|\Pi_0\big)}$ vanish when we trace $|\tilde{\beta}\>$ over $\bar{A}$.
\end{lemma}
Set $N := \tr(\Pi_0|\tilde{\Psi}\>\<\tilde{\Psi}|\Pi_0)$ and define the operator
\begin{equation}
\sigma := \tr_{\bar{A}}(\Pi_0|\psi\>\<\psi|\Pi_0) =
\frac{1}{D}\sum_{a}(|c_a^+|^2 e^{-\beta E_a^+} +|c_a^-|^2e^{-\beta E_a^-})|\psi_a\>\<\psi_a|\,.
\end{equation} 
 We can express $N = \tr(\sigma) + \tr(\Pi_0|\xi\>\<\xi|\Pi_0)$. Since 
$\tr(\Pi_0|\xi\>\<\xi|\Pi_0)= \<\xi|\Pi_0|\xi\> \le \|\xi\|^2 \le \epsilon_\fail$, 
by (\ref{eqn:proberrorbound}) we can bound $N$ and $\sigma$'s trace as follows
\begin{equation}\label{eqn:overlapbound}
(1-\epsilon_\fail) \, \frac{\cZ_\beta}{D} \, e^{-\beta\epsilon_\prec}  \le 
\tr(\sigma) \le
\frac{\cZ_\beta}{D} \, e^{\beta\epsilon_\prec}\, ,
\end{equation}
\begin{equation}
 (1-\epsilon_\fail)\, \frac{\cZ_\beta}{D} \, e^{-\beta\epsilon_\prec} \leq  N
 <\epsilon_\fail + \frac{\cZ_\beta}{D} \, e^{\beta\epsilon_\prec}.
 \end{equation} 

Because $E_a \in [0, \frac{\pi}{4}]$, we can bound the ratio $\cZ (\beta) / D$ as follows 
\begin{equation}
1 \geq \cZ(\beta)/D \geq e^{-\beta}. 
\end{equation}
By choosing the lower bound on $N$ and the upper bound on
$\|{\tr_{\bar{A}}{(\Pi_0(|\xi\>\<\xi|)\Pi_0)}}\|_{\rm{tr}}$, we obtain 
\[
\|\frac{\tr_{\bar{A}}{(\Pi_0(|\xi\>\<\xi|)\Pi_0)}}{N}\|_{\rm{tr}} \leq
\frac{\epsilon_\fail \cdot e^{\beta
\epsilon_\prec}}{(1-\epsilon_\fail){\frac{\cZ_\beta}{D}}}
\leq  \frac{\epsilon^2 \cdot e^{\beta \epsilon_\prec}}{(1-\epsilon_\fail)}
\leq 2 \epsilon^2 e^{\beta \epsilon_\prec} \leq  \frac{\epsilon}{4} . 
\]
The last inequality is obtained because $\epsilon$ is small and 
$e^x < 1 + 2x$ for $x \in [0,1]$. The term 
      	      
\begin{equation}
\|\rho_{\rm{sim}} -\frac{\tr_{\bar{A}}{(\Pi_0(|\psi\>\<\psi|)\Pi_0)}}{N}\|_{\rm{tr}}
=\sum_{i=1}^d |\frac{e^{-\beta E_i}}{\sum_j e^{-\beta E_j}} -  \frac{\tr(\sigma)}{N}|  
\leq \frac{\epsilon}{4}
\end{equation}
is still satisfied even when examining the following two extreme cases 
\begin{equation}
            \mathrm{(I)} \mbox{ Lower bound on $N$ and upper bound on $\tr(\sigma)$:}\quad \frac{e^{\beta \epsilon_\prec}}{(1 - \epsilon_\fail)e^{-\beta \epsilon_\prec}} -1 \, ,
\end{equation}                  
\begin{equation}
		\mathrm{(II)}\mbox{ Upper bound on $N$ and lower bound on $\tr(\sigma)$: } \quad 1 - \frac{(1 - \epsilon_\fail)e^{-\beta \epsilon_\prec})}{\frac{D}{\cZ_\beta}\epsilon_\fail+e^{\beta \epsilon_\prec}} .
\end{equation} 
We know that  
\begin{equation}
\sum_{i=1}^d |\frac{e^{-\beta E_i}}{\sum_j e^{-\beta E_j}} -  \frac{\tr(\sigma)}{N}|  
\leq \max 
\left\{\frac{e^{\beta \epsilon_\prec}}{(1 - \epsilon_\fail)e^{-\beta
\epsilon_\prec}} -1 , 1 - \frac{(1 - \epsilon_\fail)e^{-\beta \epsilon_\prec})}{\frac{D}{\cZ_\beta}\epsilon_\fail+e^{\beta \epsilon_\prec}} 
\right\}.
\end{equation}
Because $1+2x > e^x$ for $\forall x \in (0,1)$, we derive
\begin{equation} 
	\frac{e^{\beta \epsilon_\prec}}{(1 - \epsilon_\fail)e^{-\beta \epsilon_\prec}} -1 
	\leq \frac{1 +\frac{\epsilon}{8}}{1 - \epsilon_\fail} -1 \leq \frac{\epsilon}{4}.
\end{equation}  
In the second case because $1+x \leq
e^x$ for $\forall x \in \mathbb{R}$ and $1 \leq D/\cZ(\beta) \leq e^{\beta}$, we have 
	\begin{equation}
    1 - \frac{(1 - \epsilon_\fail)e^{-\beta	\epsilon_\prec})}{\frac{D}{\cZ_\beta}\epsilon_\fail+e^{\beta \epsilon_\prec}}
    \leq 1 -\frac{e^{-\beta \epsilon_\prec}}{\epsilon^2 + e^{\beta \epsilon_\prec}}
	\leq 1 - \frac{1 - \frac{\epsilon}{16}}{\epsilon^2 +1} 
	\leq \frac{\epsilon}{4}
	\end{equation}
for small $\epsilon$.  \qed
\section{Conclusion}
\label{conclusions}
\begin{theorem}
Let $H$ be a Hamiltonian, $\beta$ an inverse temperature
and $\epsilon \in (0,1)$ a parameter describing the desired accuracy.
Let $U_\sim$ be the Hamiltonian simulation such that $\|U - U_\sim\| \leq \epsilon_\sim$ where
$U = \exp(2 \pi i H)$. Our algorithm prepares a state $\tilde{\rho}_\beta$ that is
$\epsilon$-close to	the thermal Gibbs state $\rho_\beta$, i.e.,
\begin{equation}\label{eq:errcriteria}
\| \rho_\beta - \tilde{\rho}_\beta \|_{\mathrm{tr}} \le
\epsilon,
\end{equation}
provided that $\epsilon_\sim \le {\epsilon^2}/({8 \beta \kappa})$,
$\epsilon_\prec =	{\epsilon}/({32\beta})$ and
$\epsilon_\fail = e^{-\beta}\epsilon^2$. The complexity of our
algorithm scales like
	\begin{equation}
    	O\Big(\sqrt{\frac{D}{\cZ_\beta}}\, \frac{\beta}{\epsilon} \, (\log
	\frac{1}{\epsilon} + \beta)\Big)
    \end{equation}
in terms of the number of invocations of the controlled-$U_\sim$ operation.

\proof
The requirements for  $\epsilon_\sim$, $\epsilon_\prec$ and $\epsilon_\fail$ are immediate by 
{\em Lemma \ref{lem:SimUniPrec}} and
{\em Lemma \ref{lem:errorbound}}. By (\ref{eqn:phaseestcomplexity}) the cost for performing one Grover iteration
scales as
	\begin{equation}
    O\Big(\frac{\beta}{\epsilon} \, (\log \frac{1}{\epsilon} + \beta)\Big).
    \end{equation}
The number of Grover iterations \cite{BBH:96} is determined by
$O(\frac{1}{\tr(\sigma)}) = O(\sqrt{\frac{D}{\cZ_\beta}})$ when using the lower bound of $\tr(\sigma)$ 
in (\ref{eqn:overlapbound}).
\end{theorem}
%
\section{Acknowledgments}
		P.~W. and C.~C. gratefully acknowledge the support of NSF grants
		CCF-0726771 and CCF-0746600.
\vspace{-0.25cm}		
	

\begin{thebibliography}{99}
    	\bibitem{BACS:07}
 		D.~Berry, G.~Ahokas, R.~Cleve and B.~Sanders, {\em Efficient Quantum Algorithms for Simulating Sparse Hamiltonians},
 		Communications in Mathematical Physics, vol.~270, pp.~359--371, 2007.
 		
 		\bibitem{BBH:96}
 		M.~Boyer, G.~Brassard, P.~Hoyer and Alain Tapp, {\em Tight Bounds on Quantum Searching},
 		Fortschritte Der Physik, vol.~46(4-5), pp.~493 -- 505, 1998. 
 		
 		\bibitem{FG:99}
 		C.~Fuchs and J.~Graaf, {\em Cryptographic Distinguishability Measures for
 		Quantum Mechanical States}, IEEE Transactions on Information Theory,
 		vol.~45,  issue~4, pp.~1216--1227, 1999.
 		
 		\bibitem{JVV:86}
		M.~Jerrum, L.~Valiant and V.~Vazirani, {\em Random Generation of Combinatorial
		Structures from a Uniform Distribution}, Theoretical Computer Science,
		vol.~43, issue 2-3, pp.~169--188, 1986.
		
 		\bibitem{KLM:07}
		P.~Kaye, R.~Laflamme, and M.~Mosca, {\em An Introduction to Quantum Computing}, Cambridge University Press, 2007.
 		
 		\bibitem{Lloyd:96}
 		S. ~Lloyd, {\em Universal Quantum Simulators}, Science, vol.~273. no.~5278,	pp.~1073 -- 1078, 1996. 
 		
 		\bibitem{Metropolis:53}
		N.~Metropolis, A.~W.~Rosenbluth, M.~N.~Rosenbluth, A.~H.~Teller, E.~Teller, {\em Equation of State Calculations by Fast Computing Machines}, J.~Chem.~Phys., vol.~21, pp.~1087--1092, 1953.
	
		\bibitem{NWZ:09}
		D.~Nagaj, P.~Wocjan, Y.~Zhang, {\em Fast QMA Amplification},
		QIC vol.~9 no.~11\&12 pp.~1053--1068, 2009.
		
 		\bibitem{NC:00}
		M.~Nielsen and I.~Chuang, {\em Quantum Computation and Quantum
		Information}, Cambridge University Press, 2000.
			
		
    	\bibitem{PW:09}
 		D.~Poulin and P.~Wocjan,{\em Thermalizing Quantum Systems and Evaluating
 		Partition Functions with a Quantum Computer}, Physical Review Letter, vol.~103, pp.~220502, 2009.
 		
 		\bibitem{Suzuki:88}
 		M.~Suzuki, ed. {\em Quantum Monte Carlo Methods in Equilibrium and Nonequilibrium Systems}, 
 		vol. 74 of {\em Springer Series in Solid-State Science}, Springer, 1988.
 		
 		\bibitem{TOVPV:09}
 		K.~Temme, T.~Osborne, K.~Vollbrecht, D.~Poulin and F.~VerstraeteK,
 		{\em Quantum Metropolis Sampling},	arXiv: abs/0911.3635, 2009. 
 		
 		\bibitem{TD:00}
 		B. Terhal and D. ~DiVincenzo, {\em Problem of equilibration and the
 		Computation of Correlation Functions on a Quantum Computer}, Physical Review
 		A, vol.~61, pp.~022301, 2000.
 		
		\bibitem{WCNA:09}
 		P.~Wocjan, C.~Chiang, D.~Nagaj and A.~Abeyesinghe, {\em A Quantum Algorithm
 		for Approximating Partition Functions}, Physical Review A, vol.~80,
 		pp.~022340, 2009. 				
 		
 		\bibitem{Zalka:98}
 		C. Zalka, Proc. R. Soc. London, Ser. A, {\em Simulating Quantum Systems on a
 		Quantum Computer}, vol.~454, no.~1969, pp.~313 -- 322, 1998.
 		 		 
	
    \end{thebibliography}
 \end{document}